# A Simulation Methodology Testbed for Typhoon Sensitivity Analysis: Framework Development and Perturbation–Response Experiments with the Pangu Weather Model

Yuehua Peng[a,b], Yuchen Zhang[c], Qin Huang[d], Chengzhi Ye[b], Jingsong Yang[a]

[a] State Key Laboratory of Satellite Ocean Environment Dynamics, Second Institute of Oceanography, Ministry of Natural Resources, Hangzhou, 310012, China

[b] Hunan Institute of Meteorological Sciences, Hunan Meteorological Bureau, China Meteorological Administration, Changsha 410118, China

[c] School of Aerospace Engineering, Tsinghua University, Beijing 100084, China

[d] School of Complex Adaptive Systems & Water Institute, Arizona State University, Tempe, AZ, USA

*Corresponding author*: Peng Yuehua, pengyuehua@hotmail.com

## Abstract:

Understanding how typhoons respond to localized perturbations in their environmental fields is fundamental to assessing the limits of predictability and exploring the potential for influencing track or intensity intervention. This study develops a dedicated simulation methodology testbed for typhoon sensitivity analysis by integrating the Pangu weather model—a high-precision AI forecasting system—with Proportional-Integral-Derivative (PID) closed-loop techniques. The testbed is constructed with modular functional blocks including a meteorological prediction module, an artificial perturbation input interface, a typhoon quantitative modeling module, and a PID closed-loop test module, implemented via a cross-platform MATLAB/ONNX technical framework. A Single-Input Single-Output (SISO) test system was built, with velocity and thermal perturbations set as the core inputs and typhoon track and intensity as the key output targets, to perform controlled perturbation–response experiments. The experiments reveal the feasible perturbation–response range, the parameter tuning behavior of the PID module, and the energy-scale response characteristics under different perturbation modes, and quantify the input–output coupling relationships of the test system. By constructing this testbed on an operational AI weather forecasting model, this study provides a framework that goes beyond idealized sensitivity studies typically validated only on low-order dynamical models. The testbed offers an expandable platform for investigating typhoon sensitivity to artificial environmental perturbations and provides a foundation for subsequent expansion toward multi-input multi-output architectures and advanced analysis strategies such as nonlinear PID or model predictive control.

**Keywords** : Pangu weather model, typhoon sensitivity analysis, PID control, SISO test system, artificial environmental perturbations

## 1 Introduction

Against the backdrop of global climate change, the intensification of typhoon activity has raised urgent research needs for a deeper understanding of how these storms respond to changes in their environment, and simulation-based sensitivity analysis has become a crucial approach in advancing this field. However, current sensitivity studies on typhoon response to perturbations face the problem of lacking a standardized simulation methodology testbed. Most perturbation experiments are only verified on idealized simplified models or traditional numerical weather models, and there is a lack of test platforms built on large-scale operational AI weather forecasting models that offer high prediction accuracy and fast inference speed. As a mesoscale meteorological phenomenon, the sensitivity of typhoons involves complex nonlinear dynamical systems, and the construction of a specialized testbed is the basic premise for systematic perturbation–response investigation, performance characterization, and method optimization.

Prior research includes the Ross Hoffman team's theoretical framework based on nonlinear optimal control technology, with simulation tests on the MM5-4DVar mesoscale system (2004, 2006). Subsequently, Peng Yuehua, Wang Ting, and other scholars modified the WRF model's 4DVar system and introduced vortex dynamic initialization to simulate sensitive perturbation responses of tropical cyclone track (Wang et al., 2021; Peng et al., 2023). However, these efforts lack standardized testbed support, and perturbation results are difficult to apply directly to actual AI meteorological models. Japan's MS8 program has carried out research on EnMPC and other control methods for weather intervention, but its core verification remains based on the idealized Lorenz 63 model, with no testbed built on an actual large-scale AI weather forecasting model. This creates a large gap between simulation-based sensitivity analysis and real-world application scenarios. For complex and high-dimensional AI weather forecasting systems, the PID method, with its simple structure, strong robustness, and easy module integration, provides an ideal foundational framework for initial testbed construction and serves as a modular basis for subsequent integration of more sophisticated perturbation analysis methods.

The Pangu Weather Large Model, published in *Nature*, is the first AI weather forecasting method that surpasses the accuracy of traditional numerical forecasting methods, with ultra-fast local inference speed (approximately 1.4 seconds) and high prediction accuracy for typhoon track and intensity. Its excellent performance makes it the core meteorological prediction module of the typhoon sensitivity analysis testbed and provides necessary technical support for real-time iterative perturbation–response experiments.

Based on the above research gaps and technical advantages, this study constructs a simulation methodology testbed for typhoon sensitivity analysis with the Pangu Weather Large Model as the core prediction module and PID closed-loop techniques as the foundational experimental module. We first complete precision testing and performance characterization of the Pangu model as the testbed's prediction module, then develop an artificial perturbation input interface and design a typhoon quantitative modeling module adapted to the SISO test system. We further build a complete testbed technical implementation framework based on MATLAB and ONNX and construct a SISO test system to carry out qualitative perturbation–response experiments under velocity and thermal forcings. Finally, by comparing with the MS8 program, we summarize the testbed's advantages and current limitations and propose module expansion and methodology optimization directions. This study aims to provide a standardized, reusable, and expandable simulation methodology testbed for typhoon sensitivity analysis research.

This study adopts the Pangu Weather Large Model primarily due to its high computational speed and exceptional typhoon prediction accuracy, as detailed in Section 2. Section 3 examines the perturbation input framework and typhoon modeling requirements. Section 4 elaborates on the technical roadmap for typhoon sensitivity experiments, including the developed simulation platform. Section 5 presents qualitative perturbation–response experiments for the SISO system, with particular emphasis on velocity and temperature perturbation cases. Section 6 compares this research with Japan's MS8 program and proposes key research priorities for future studies.

## 2 Pangu Weather Model and Its Typhoon Prediction Accuracy

### 2.1 Pangu Weather Model

The Pangu weather large model, first proposed by researchers from Huawei Cloud, was published in <Nature> on July 6, 2023. This model is the first AI method to surpass the accuracy of traditional numerical forecasting methods, with prediction accuracies from 1 hour to 7 days all exceeding those of the European Centre for Atmospheric Research's operational IFS. Its local inference time is approximately 1.4 seconds, which is over 10,000 times faster than traditional numerical methods. This forms the basis for applying cybernetics to actual typhoon regulation, since closed-loop iterative optimization using traditional numerical models would be infeasible due to time costs. Therefore, this study uses AI weather forecasting models instead of numerical models.

The model employs a specialized architecture termed the 3D Earth-specific transformer. It integrates 13-layer atmospheric data, combining upper-air and surface information variables, into a deep neural network, which generates a 3D cube through various downsampling techniques. These data are then processed via an encoder-decoder architecture to predict future conditions. Inspired by the Swin transformer, this structure represents a variant of the Vision transformer, with the decoder section employing up-sampling interpolation to restore original resolution. Each 3D deep network contains approximately 64 million parameters.

Based on this architecture, Bi et al. (2023) selected 88 tropical cyclones occurring in 2018 for comparison of numerical accuracy and direct positional error at the eye of the storm. As shown in Figure 1, the Pangu model demonstrates clear advantage over traditional methods for extreme climate prediction. For the most important path prediction in this paper, results of two super typhoons in the western Pacific, Connie and Yutu, were also used for comparison. The model has very high accuracy in forecasting typhoon paths in the western Pacific, which fits well with this paper's focus.

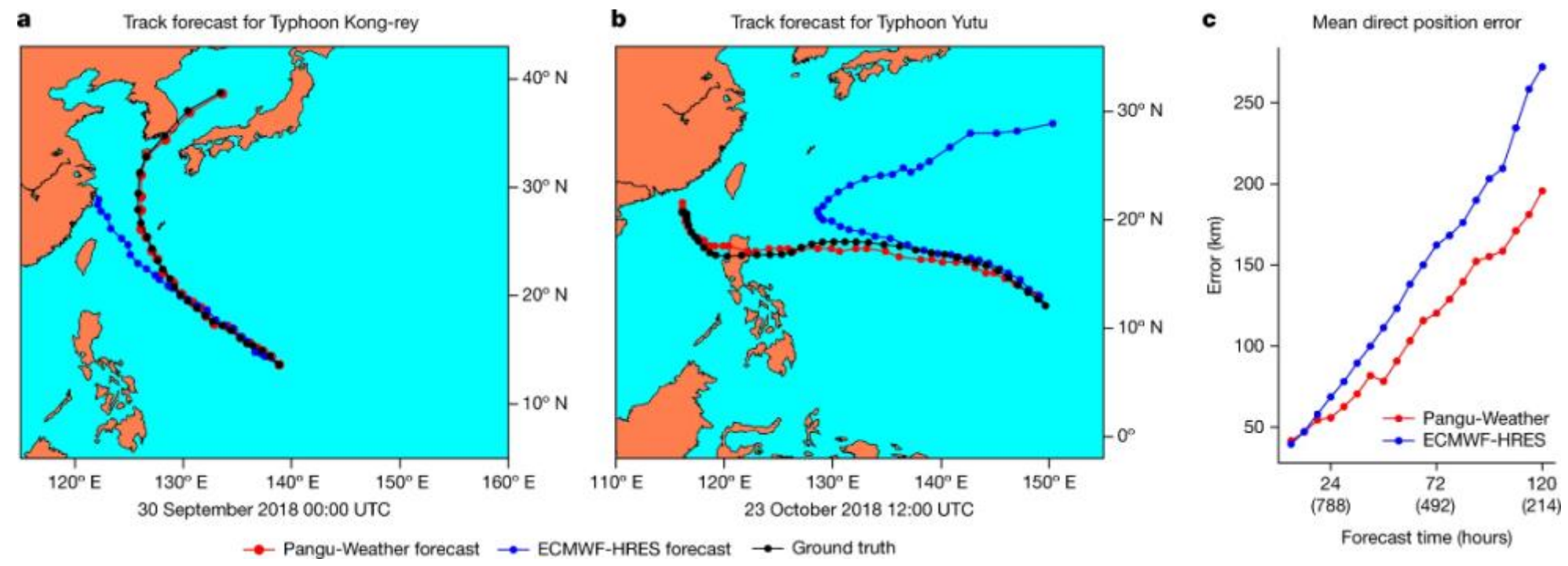

Figure 1. Comparison of Pangu weather model predictions with ECMWF-HRES (from Bi et al., 2023)

## 2.2 Error Analysis of Typhoon Prediction

The PanGu weather supermodel utilizes data from the European Centre for Medium-Range Weather Forecasts (ECMWF), which are globally interpolated and corrected, representing the most widely used format in meteorological forecasting. This enables direct comparison between ECMWF raw data and the supermodel's predictions regarding typhoon forecast errors. Error analysis focuses specifically on typhoon and surrounding area predictions.

Using the Pangu Weather Large Model, meteorological data from September 30, 2018 at 00:00 was applied to predict the 6-hour weather field in the Pacific Ocean. At this time, the typhoon was primarily located in the local Pacific region, with coordinates ranging approximately from 137°E to 150°E and 7°N to 18°N. To validate prediction accuracy, the actual pressure field from ECMWF at the same location was employed. The two-dimensional surface pressure field was calculated using relative error analysis, with the error distribution shown in Figure 2. Calculations revealed that the Pangu model's predicted pressure field in the local Pacific region exhibited errors below 0.1% in most areas. While some specific regions around the typhoon eye showed slightly higher errors, the overall error remained below 0.3%, indicating high precision in pressure prediction.

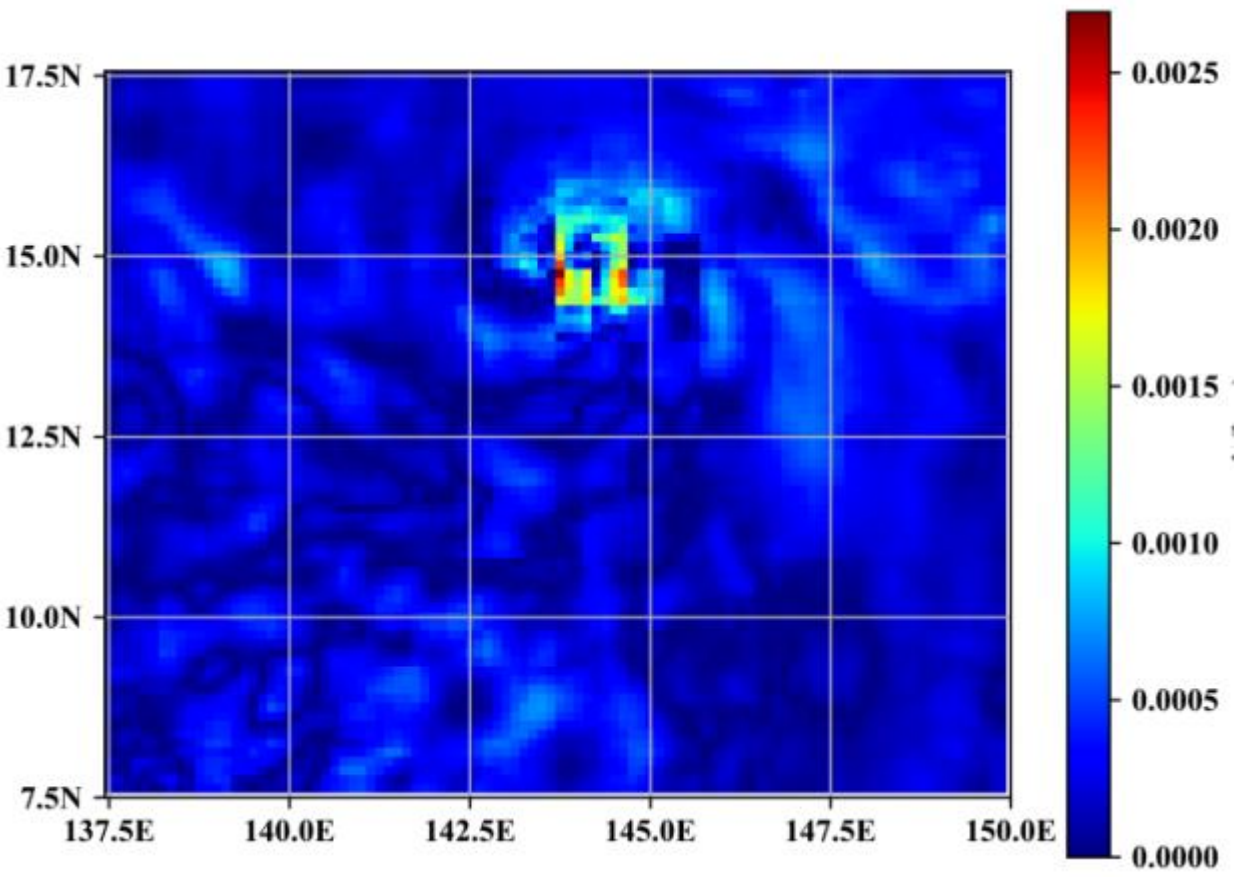

Figure 2. Distribution of pressure relative errors between Pangu and ECMWF at 06:00 on September 30, 2018

Similarly, we compared the local Pacific wind speed field predicted by the Pangu model for the six hours after 00:00 on September 30, 2018, with the actual measured wind speed field from ECMWF at 06:00 on the same day. The comparison reveals that while the overall trend of wind speed distribution is consistent, direct numerical error analysis using full physical fields is insufficient due to gradient-induced high wind speeds around typhoons. We therefore employ spatial averaging for comparison. Calculations show the wind speed error is approximately 0.3%.

For the temperature field, Figure 3 presents a spatial distribution map of error percentages, calculated as the percentage difference between predicted and actual values. The analysis reveals that calculated errors are predominantly concentrated in land areas, while oceanic regions show temperature deviations consistently below 1%. These findings conclusively demonstrate the model's high accuracy and reliability in weather forecasting.

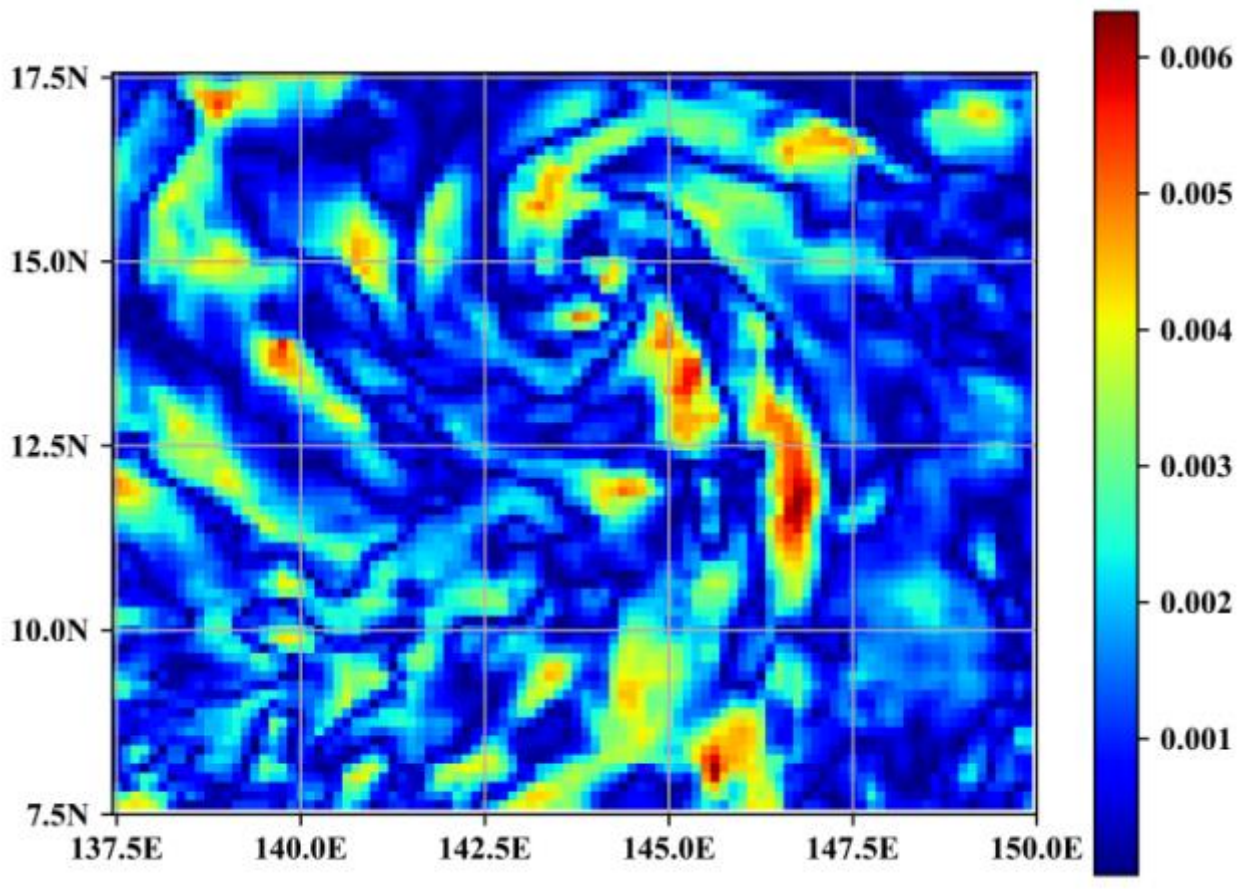

Figure 3. Distribution of temperature relative errors around the typhoon calculated by the model

# 3 Artificial Perturbation Framework and PID-Controlled Response Experiments

## 3.1 Artificial Perturbation Addition

The Pangu weather supermodel has achieved prediction of extreme weather events, but this study aims to achieve artificial typhoon regulation through cybernetics. Since the original Pangu model cannot meet this objective, modifications are required to provide programming interfaces for human-induced disturbances and to validate the supermodel's feedback mechanisms against such interventions.

The PanGu large-scale model utilizes training data from ECMWF, provided in .nc file format. After the model processes raw data, an additional data processing layer is added to expose an interface, enabling external access for manual adjustments to meteorological data. This approach ultimately achieves artificial regulation.

Artificial regulation of data without closed-loop control can be understood as open-loop control for typhoon regulation, serving as fundamental validation. Extreme artificial interference was used to validate system viability. In feasibility validation, artificial disturbances at 100 m/s were applied to predict typhoon trajectories. The original data was derived from Typhoon Connie on September 30, 2018. As shown in Figure 4, the arrow indicates the disturbance direction. The red line represents the original trajectory, while the blue line shows the trajectory feedback after disturbance.

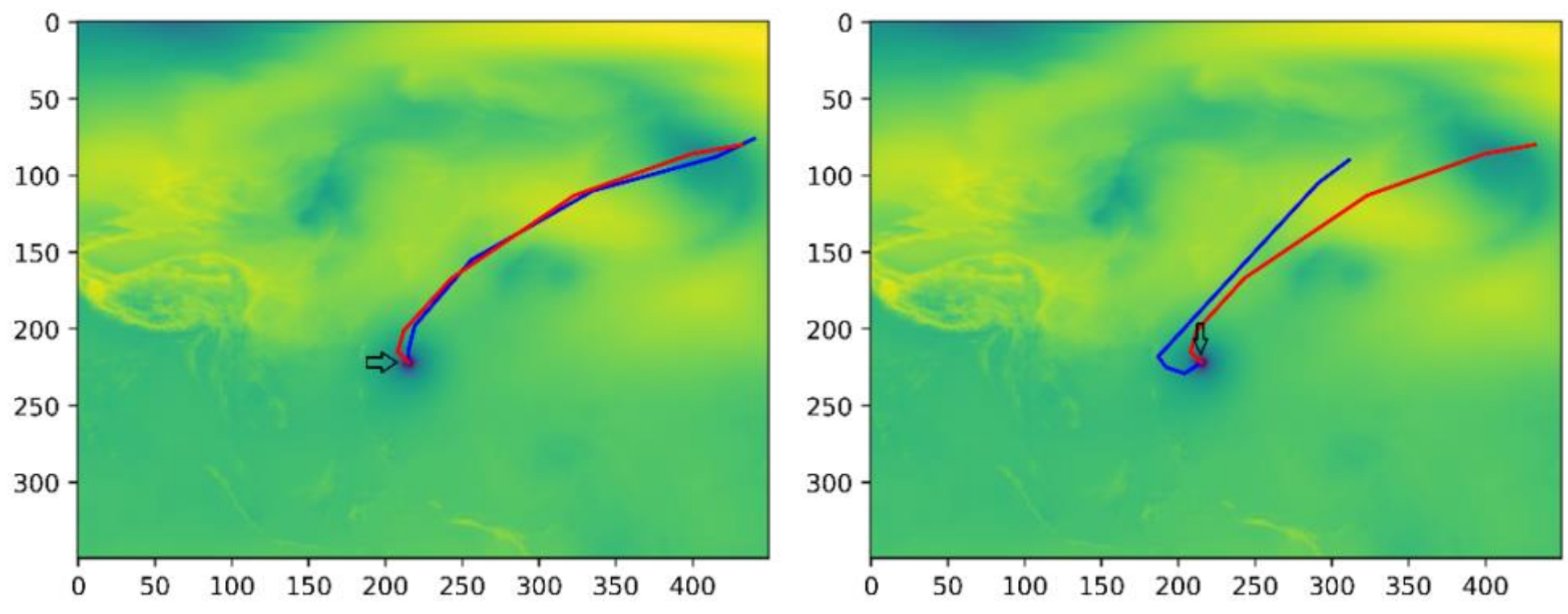

Figure 4 Long-term open-loop prediction results based on artificial interference

The results clearly demonstrate the model's effective response to artificial interference. During the initial six-hour phase, artificial interference significantly altered the typhoon's path. However, subsequent hours saw the trajectory gradually revert to its original course. The right-hand figure demonstrates that after interference application, the typhoon's path completely deviated from its original trajectory. This phenomenon arises from the highly nonlinear nature of typhoons, influenced by global meteorological factors. Local interventions exert nonlinear effects, making long-term typhoon modification inherently unpredictable. In contrast, short-term control measures remain relatively manageable, particularly when the typhoon is near the interference zone. Therefore, this study focuses on short-term typhoon modification.

### 3.2 Introduction of PID

The proportional (P), integral (I), and derivative (D) components of a PID controller correspond to current error, accumulated error, and future error, respectively. When system characteristics are unknown, the PID controller is generally considered the most suitable choice. Given the unknown characteristics of the Pangu weather system, employing a PID controller is the most natural option.

PID controllers can be categorized into traditional single-input single-output (SISO) systems and multi-input multi-output (MIMO) systems based on multi-loop control or fuzzy control theory. Since the focus of this study is on applying cybernetics principles to practical typhoon regulation rather than control methodology itself, the simplest SISO system is employed. Due to SISO constraints, independent and dependent variables are restricted to scalar values for robustness. Four combinations of control methods were selected, as shown in Table 1.

Table 1 Closed-loop control implemented in this study

| COMPOSITE MARKING | ARGUMENT | DEPENDENT VARIABLE |
|---|---|---|
| **COMBINATION 1** | velocity perturbation | Typhoon track |
| **COMBINATION 2** | velocity perturbation | typhoon intensity |
| **COMBINATION 3** | thermal perturbation | Typhoon track |
| **COMBINATION 4** | thermal perturbation | typhoon intensity |

This leads to typhoon modeling. Unlike velocity and temperature disturbances, which can be directly applied as intuitive physical quantities, dependent variables are relatively abstract concepts. Both trajectory and intensity are critical targets for typhoon modification. Controlling trajectories can direct impact to less affected areas, while regulating intensity mitigates unavoidable impacts.

### 3.3 Modeling of Typhoons

Typhoon movement trajectories and intensity are abstract concepts and cannot be directly used as control targets. Rigorous modeling is essential. Based on SISO assumptions, the modeling objective is defined as a directly usable scalar quantity.

As shown in Figure 5, the Pacific surface pressure field of Typhoon Connie reveals an exceptionally complex structure. The eye of the typhoon is not an abstract point but a vast region, and its rotating influence extends over an enormous surrounding area. Abstracting a directly usable scalar remains challenging.

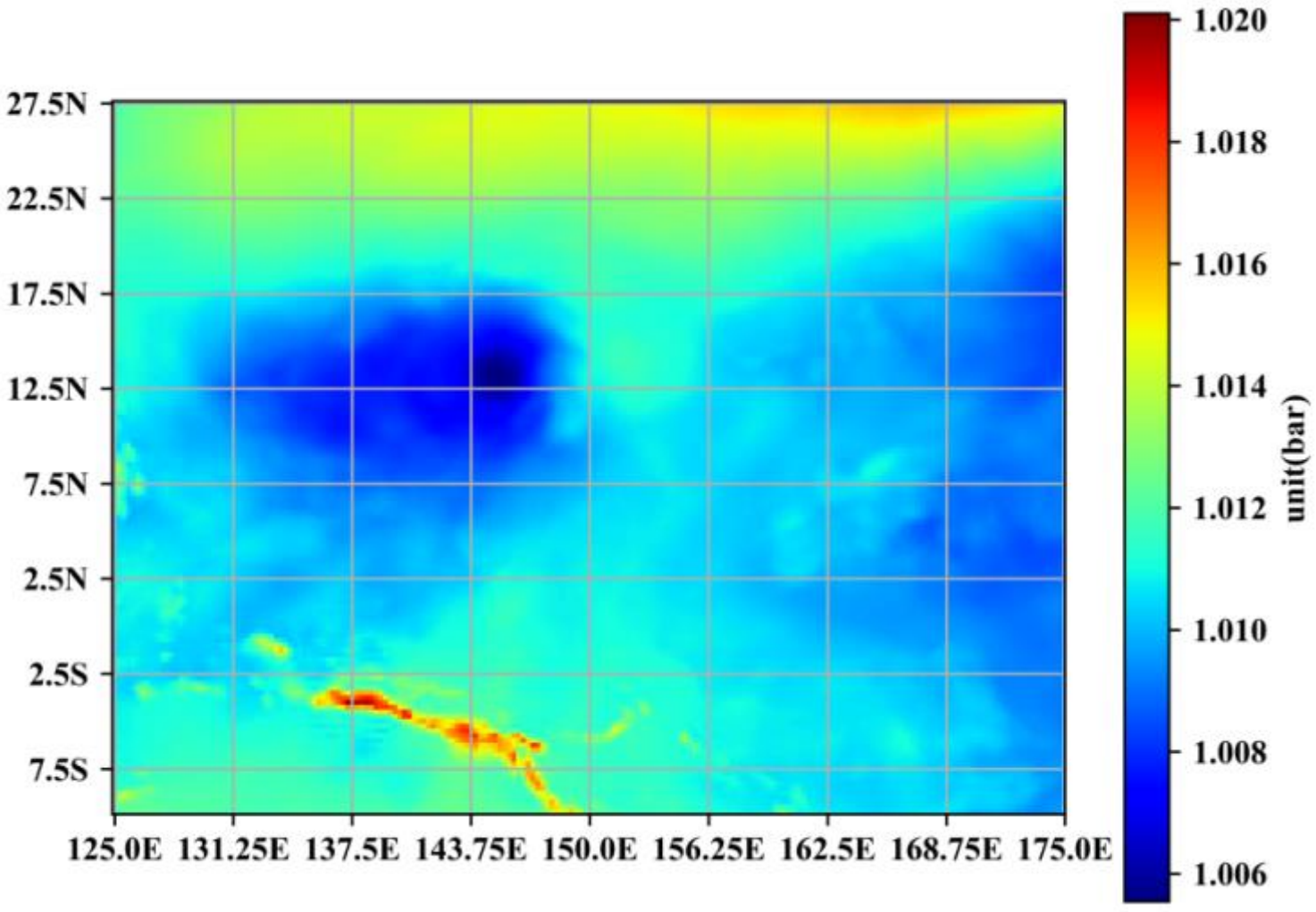

Figure 5 Surface pressure field of Typhoon Connie in the Pacific Ocean on September 30, 2018

Unlike Figure 5, Figure 6 presents a magnified view of the typhoon eye. The analysis reveals that the typhoon eye exhibits a more regular circular structure, accompanied by a distinct pressure gradient that abruptly changes relative to surrounding regions.

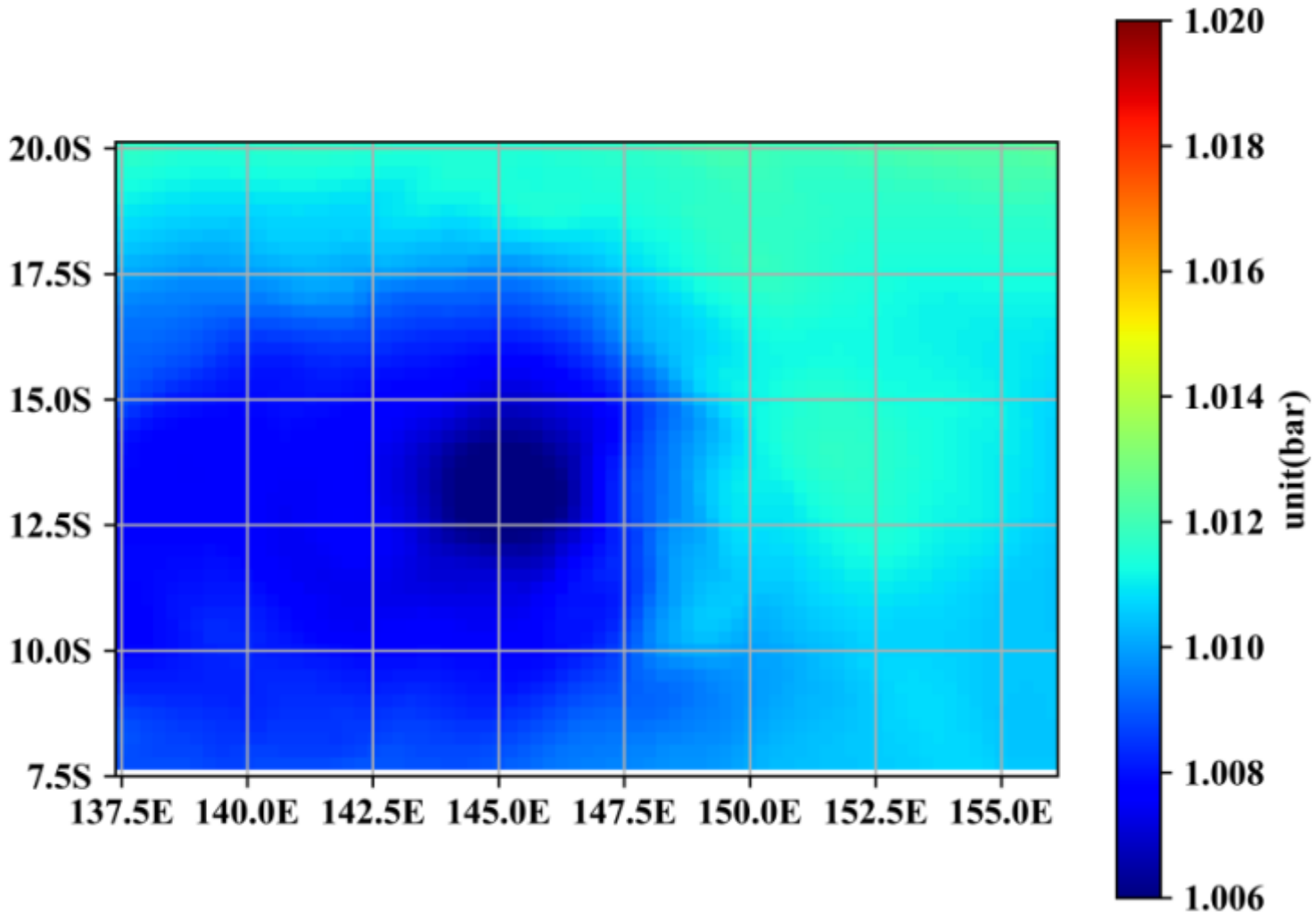

Figure 6 Surface pressure field of Typhoon Connie in partial enlargement on September 30, 2018

Based on this observation, modeling can focus on the eye. The trajectory of the typhoon can be abstracted as the trajectory of its eye, and intensity as properties of the eye. Two approaches exist for determining trajectory: geometric center or local minimum pressure point. The second method, using the lowest local pressure point, proves most straightforward. This localized minimum maintains strong correlation with the geometric center and represents the region of maximum intensity. Consequently, this project adopts this modeling approach.

Considering the scalar assumption for PID control robustness, the displacement vector of the typhoon cannot be satisfied. Therefore, further modeling of displacement is conducted via polar coordinate transformation:

$$r=\sqrt{x^2+y^2}$$

$$\theta = \arctan\left(\frac{y}{x}\right)$$

The polar coordinate system provides an ideal framework for expressing angular relationships and spatial distances between two points. By converting vector displacement into motion angles and distances, we achieve precise tracking. The typhoon's motion angle serves as the key output parameter for trajectory modeling. Typhoon intensity is modeled using the pressure value corresponding to the local minimum point in the eye, providing a scalar characteristic of overall intensity.

# 4 Technical Route and Simulation Platform for Typhoon Sensitivity Experiments

## 4.1 Technical Approach for Typhoon Sensitivity Experiments

Based on the aforementioned analysis, precise prediction of typhoon dynamics can be achieved using initial data. By applying mechanisms to alter individual variables, energy distribution of the typhoon and surrounding wind field is modified, enabling controlled deviations from predetermined trajectories and facilitating directional intensification or weakening. Employing the PID control method with closed-loop principles enables targeted artificial manipulation of typhoons.

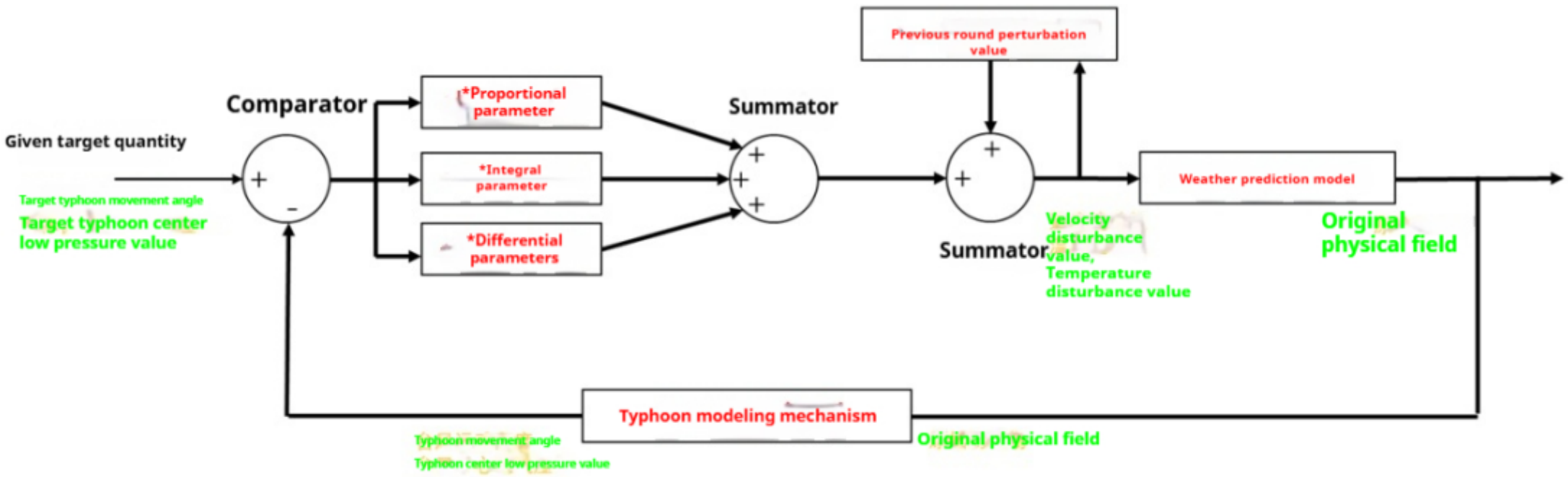

Figure 7. Closed-loop control of directional artificial regulation of typhoon based on proportional-integral-differential method

## 4.2 Code Implementation for MATLAB Platform

The specific interface of the Pangu weather model is Open Neural Network Exchange (ONNX). ONNX addresses the challenge that neural network models trained on one framework cannot be directly deployed on another. It supports numerous machine learning platforms including PyTorch and Caffe2. MATLAB provides the importNetworkFromONNX() function to interface with ONNX networks. However, the training platform version of the large model we use is based on ONNX version 14 (GPU-accelerated) and 12 (CPU-solved). Since MATLAB 2023b ONNX interface functions only support version 7, direct use of MATLAB's network interface functions cannot achieve the intended functionality.

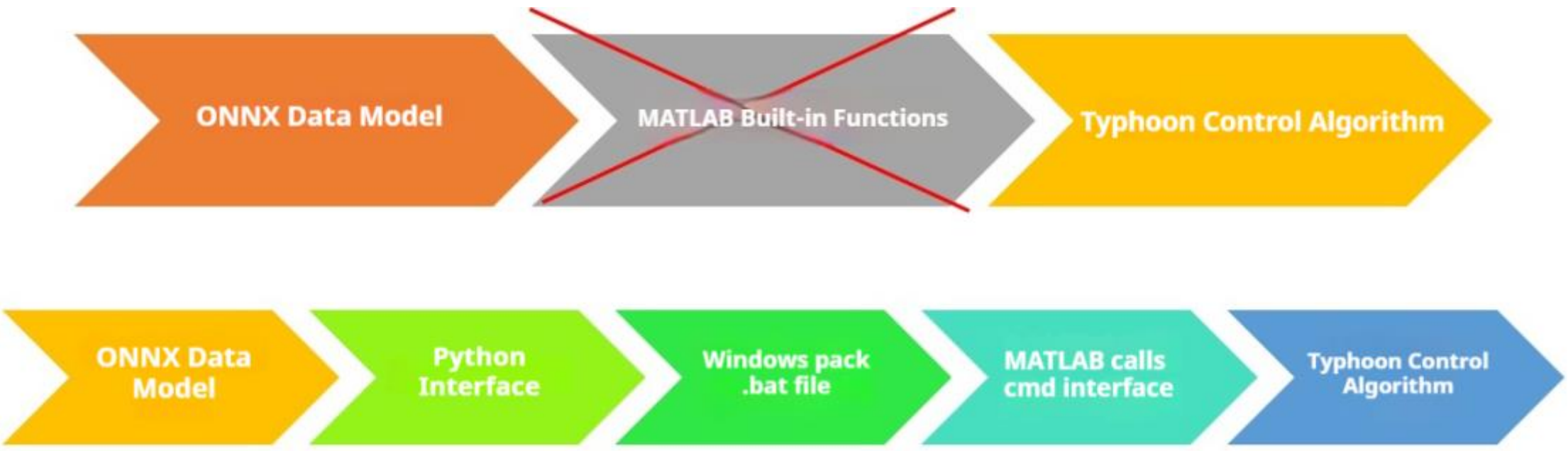

Figure 8. Technical roadmap for MATLAB-based code implementation.

As illustrated in Figure 8, given ONNX's superior Python support, leveraging Python and Windows command-line tools as an intermediary proves intuitive and efficient for executing MATLAB platform code. The command-line approach also automatically releases memory after program execution, which is particularly advantageous for closed-loop control processes.

### 4.3 Simulation Platform for Typhoon Sensitivity Experiments Based on Cybernetics

A simulation platform for artificial typhoon regulation was developed using the PID control method. This project abstracted the complex typhoon prediction model into multiple controllable SISO systems. System inputs were defined as intensity of human-induced disturbances, specifically velocity, temperature, and pressure perturbations.

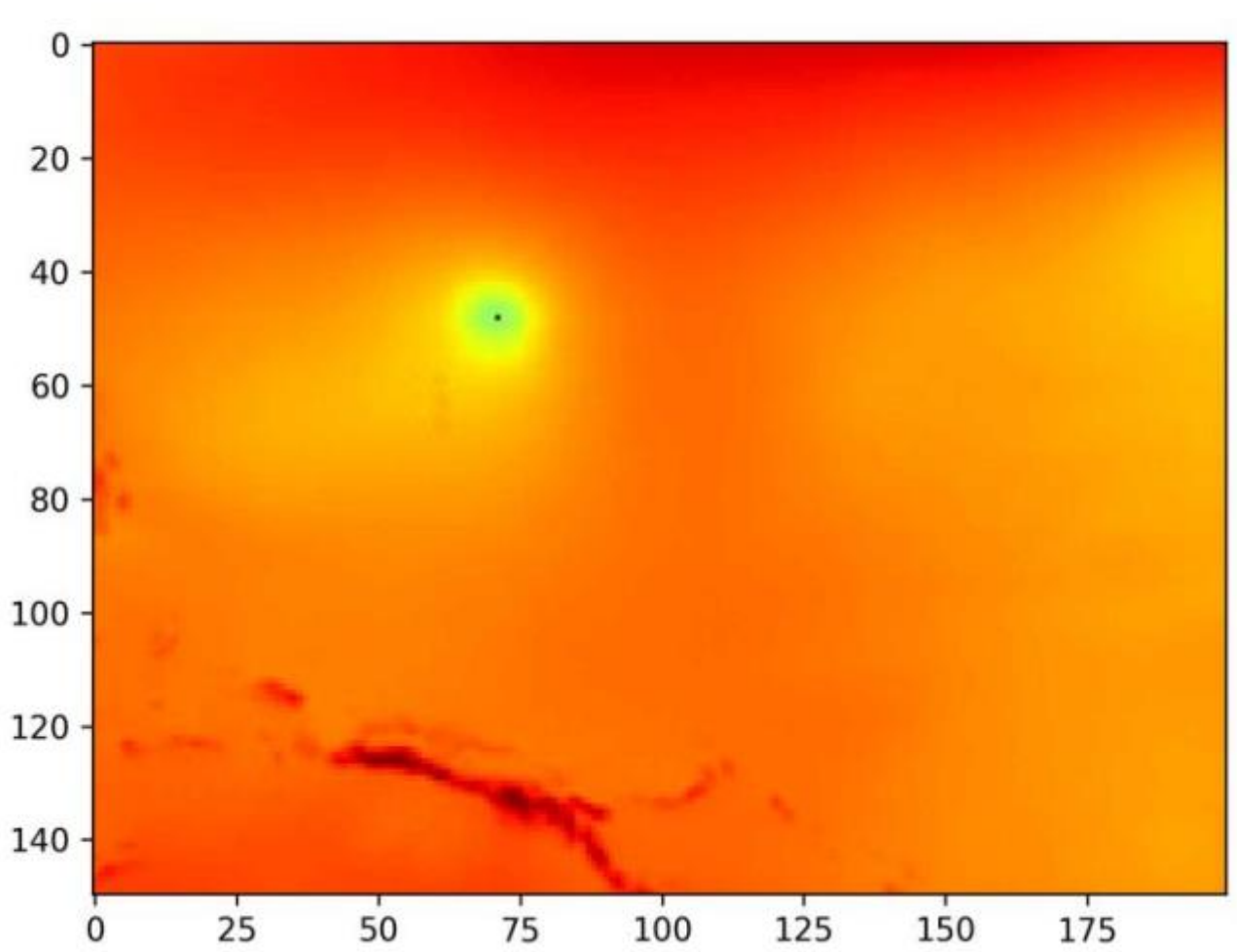

Figure 9 shows the typhoon eye's position obtained by the extrema local search algorithm.

To achieve system output, we implement typhoon modeling to quantify target control parameters. The process begins with locating typhoon positions using the Extrema Local Search algorithm, which identifies local maxima in surface pressure data. As demonstrated in Figure 9, this method has proven effective in tracking typhoon centers.

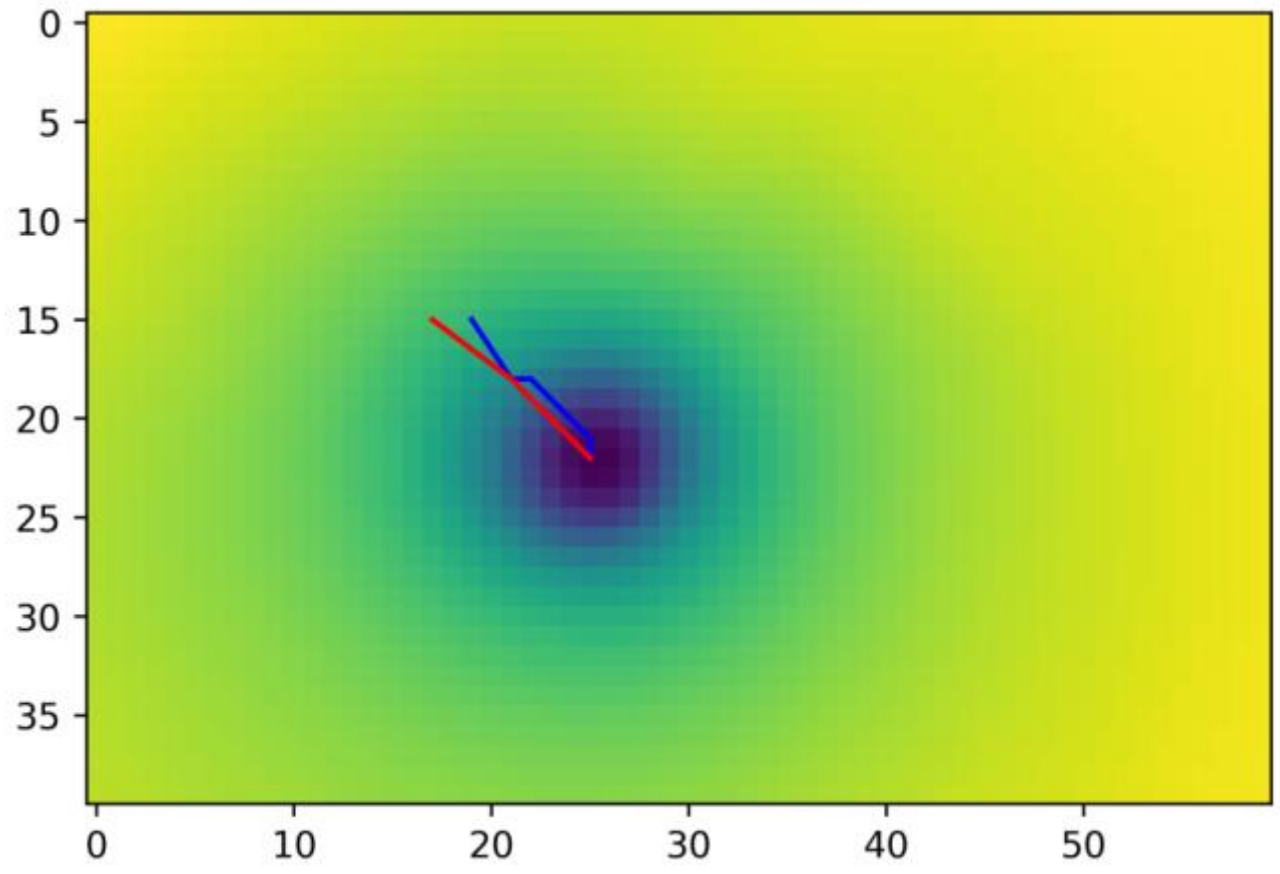

Figure 10 Schematic of local typhoon movement trajectories within 1 hour (red: original; blue: with horizontal velocity perturbation)

As shown in Figure 10, the movement trajectory is highly two-dimensional. Polar coordinate transformation yields angular parameters used as system outputs. For intensity regulation, due to resolution limitations of the Pangu model (0.25° × 0.25°), accurate quantification of typhoon size cannot be achieved without local interpolation, which compromises accuracy. Therefore, this project utilizes central air pressure value as system output.

# 5 Demonstration Case Studies: Perturbation–Response Runs of the Testbed

This section presents four example runs to demonstrate that the testbed is technically functional and can produce interpretable outputs. Each run follows a fixed protocol (feasibility range scan → setpoint selection → closed-loop PID tracking). Numerical values reported are testbed-generated outputs under specific parameter settings and should not be interpreted as real-world estimates.

## 5.1 Speed Perturbation Based Control

Since typhoon regulation operates as a highly opaque system, the first critical step is to determine the system's feasible control range. We used extreme control values (+100 and -100) to identify the adjustable range for trajectory angle, which spans from 116° to 180°. We selected 150° as the target value. By empirically adjusting PID parameters, we obtained the results in Figure 11.

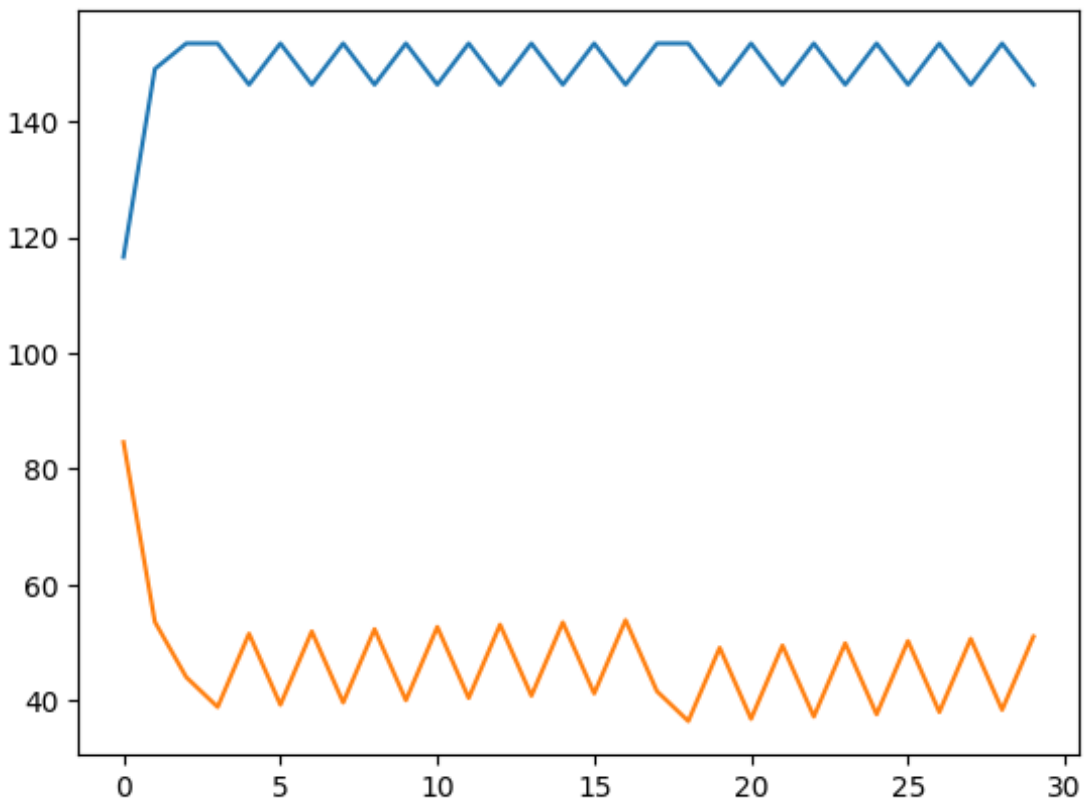

Figure 11 System output angle (blue) and input speed (orange-red) controlled by speed disturbance

Due to the 6-hour prediction model, the central movement direction is influenced by resolution limitations and does not exhibit continuous distribution. After three adjustments, the system's feedback steadily oscillates around 150°. The disturbance velocity oscillates within the 43 m/s range. According to the total energy formula

$$E=\rho*(Cv \cdot T+\frac{1}{2}u^2)$$

energy per unit volume is approximately 1,193 J. Considering the disturbance zone utilized, total energy consumption amounts to about 200 billion kWh.

The second SISO control system regulates typhoon intensity (central pressure output) via speed disturbances. The adjustable range is approximately 1.002–1.00484 bar, with target 1.0032 bar. Table 2 lists PID parameters; Figure 12 shows Test Group 1 output. The system output curve from the empirical adjustment method (Figure 13) corresponds to the PID parameters of Test Group 2-5 in the table 2.

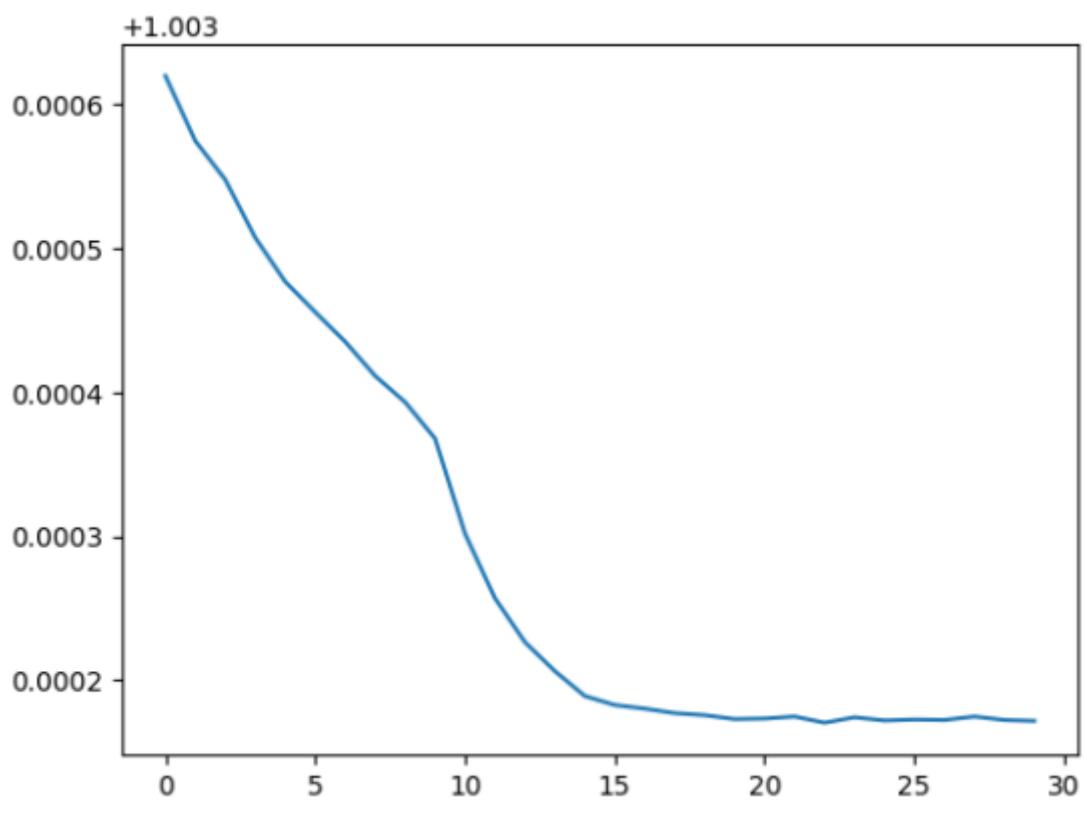

Figure 12 System output with parameter combination for Test Group 1

Table 2 Regulation of typhoon intensity based on speed disturbance

| | PROPORTIONAL PARAMETER | INTEGRAL PARAMETER | DIFFERENTIAL PARAMETER |
|---|---|---|---|

| **TEST GROUP 1** | 10000 | 100 | 10 |
|---|---|---|---|
| **TEST GROUP 2** | 10000 | 10 | 10 |
| **TEST GROUP 3** | 1000 | 100 | 10 |
| **TEST GROUP 4** | 10000 | 100 | 1 |
| **TEST GROUP 5** | 100000 | 100 | 10 |

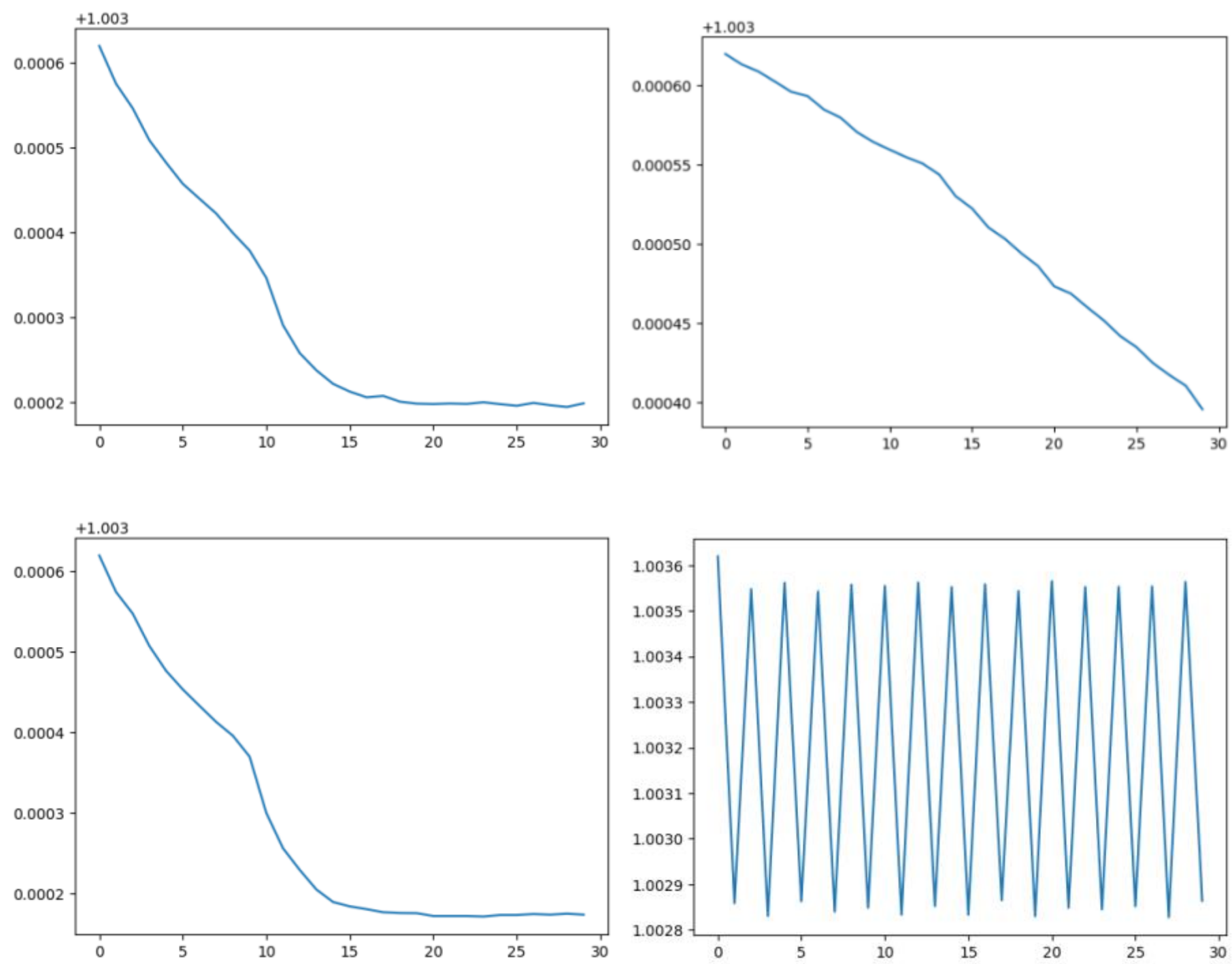

Figure 13 Study on proportional-integral-differential parameter tuning (Test Groups 2-5)

Parameter comparison reveals standard PID behavior: differential suppresses fluctuations; integral eliminates stability bias; proportional affects convergence speed and overshoot risk. These characteristics validate the feasibility of PID closed-loop control in typhoon regulation systems. The input speed converges to approximately 16.5 m/s (Figure 14). Energy per unit volume is about 176 J, total consumption about 30 billion kWh.

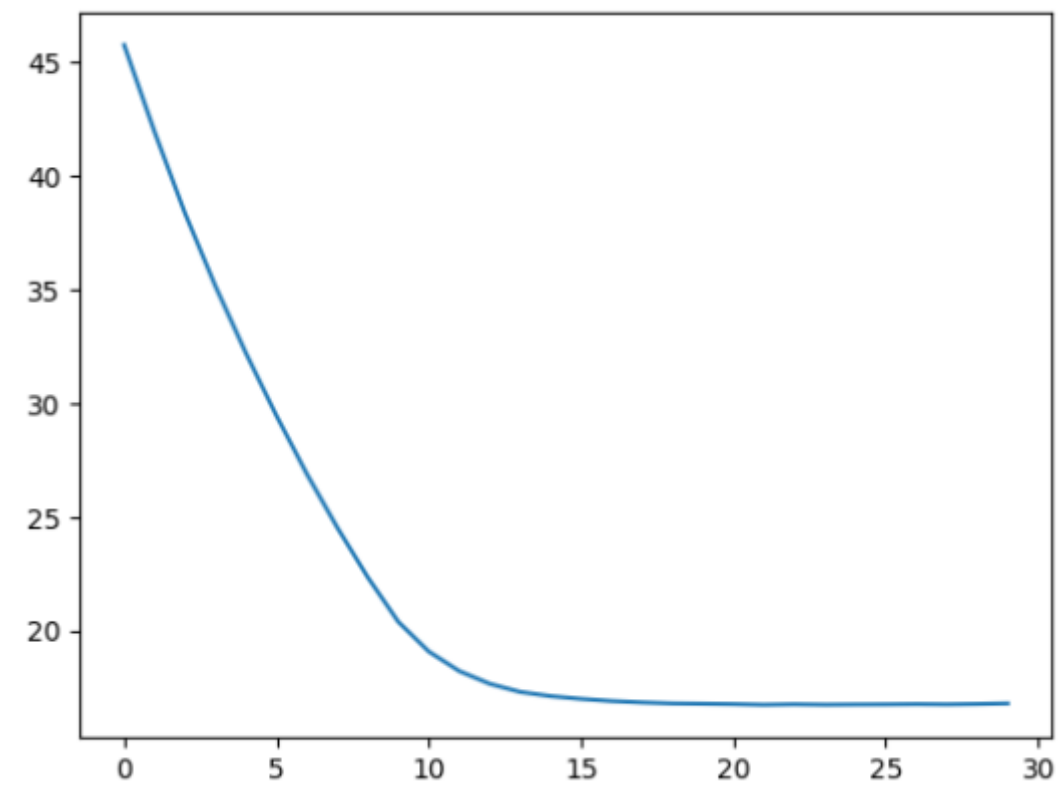

Figure 14 Closed-loop control results of system input for Test Group 1

## 5.2 Temperature Perturbation-Based Regulation

Empirical observations reveal that typhoon trajectories demonstrate minimal sensitivity to temperature fluctuations, consistent with fundamental dynamics. This section focuses on temperature disturbance-based regulation of typhoon intensity. The operational range is approximately 1.0021–1.0025 bar, with target 1.00245 bar (original 1.00236 bar).

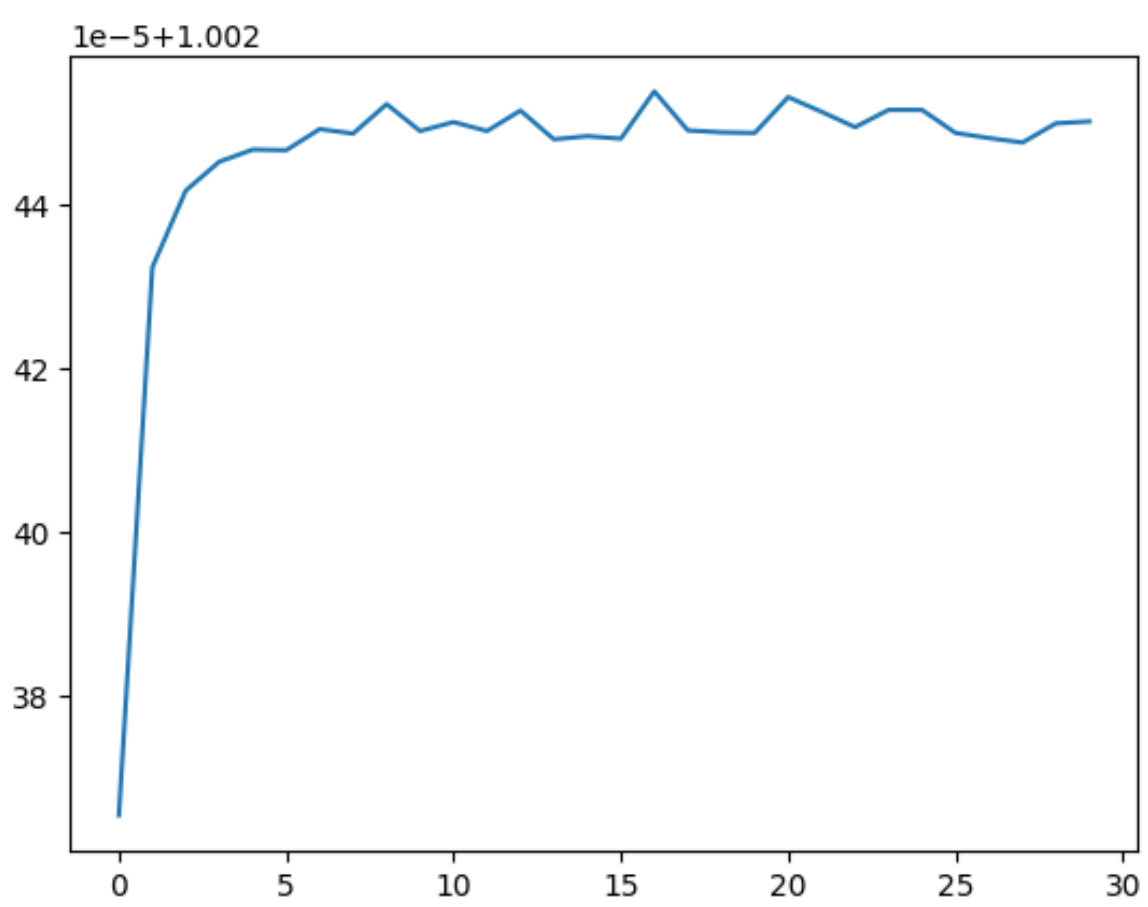

Figure 15 System output for typhoon intensity modulation using temperature perturbation

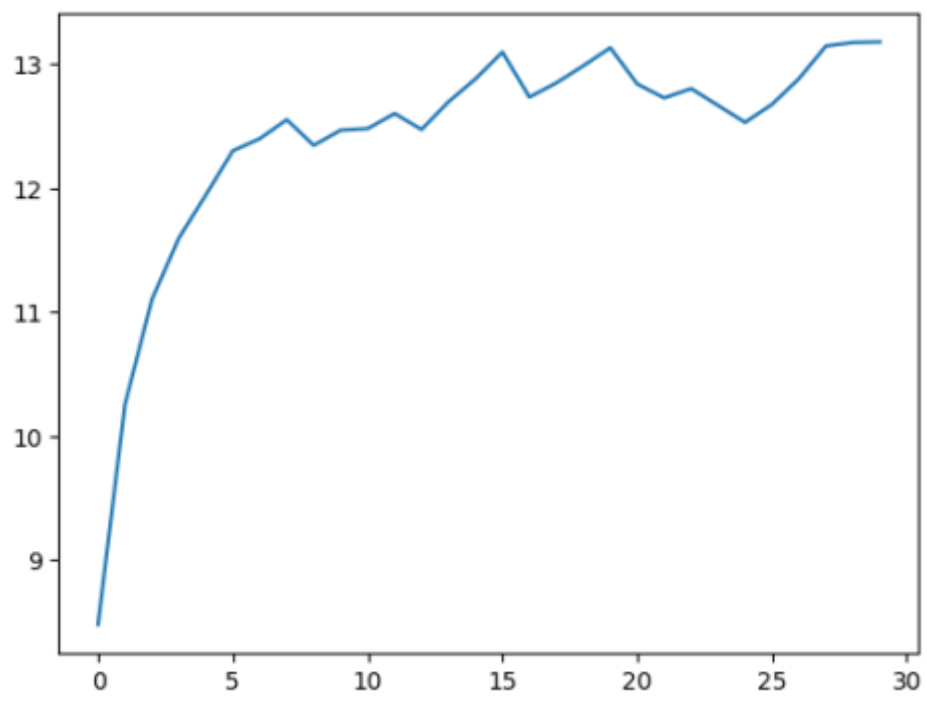

Figure 16 System input of temperature disturbance regulation on typhoon intensity

As shown in Figure 15, the final result converges to 1.00245 bar. However, due to temperature's nonlinear influence, significant oscillations persist near convergence. The input oscillates around 13 K (Figure 16). Energy per unit volume is approximately 16,770 J, with total consumption around 3 trillion kWh.

Response tests with disturbance energies at approximately 1%, 0.1%, and 0.01% of typhoon energy show that system response decreases nonlinearly with input energy magnitude (Figure 17). The Pangu model integrated with PID controller forms a functional simulation system (Figure 18).

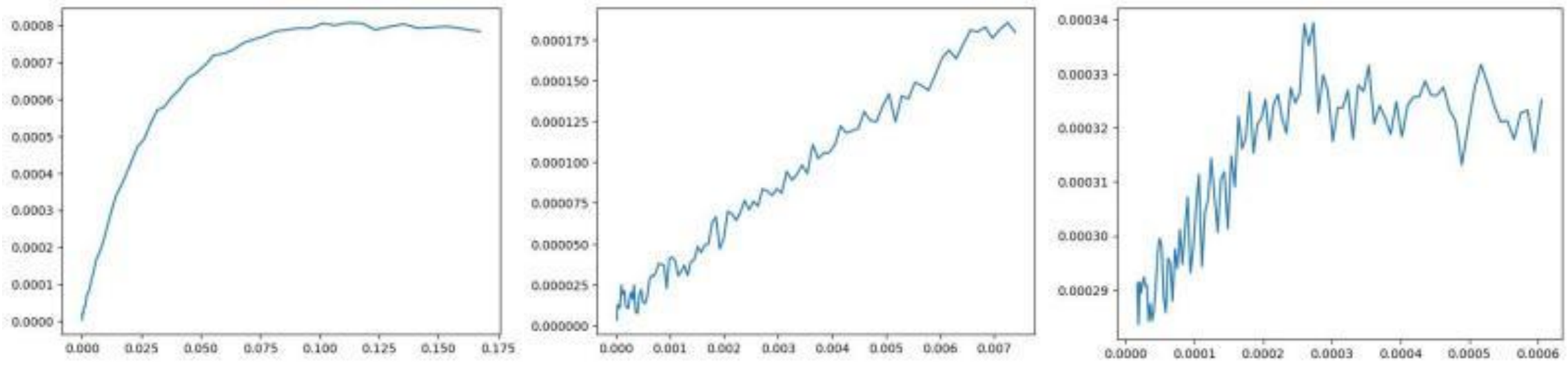

Figure 17 Variation rate of typhoon pressure extremum under different disturbance energies: (a) Energy of about 1%, (b) Energy of about 0.1%, (c) Energy of about 0.01%

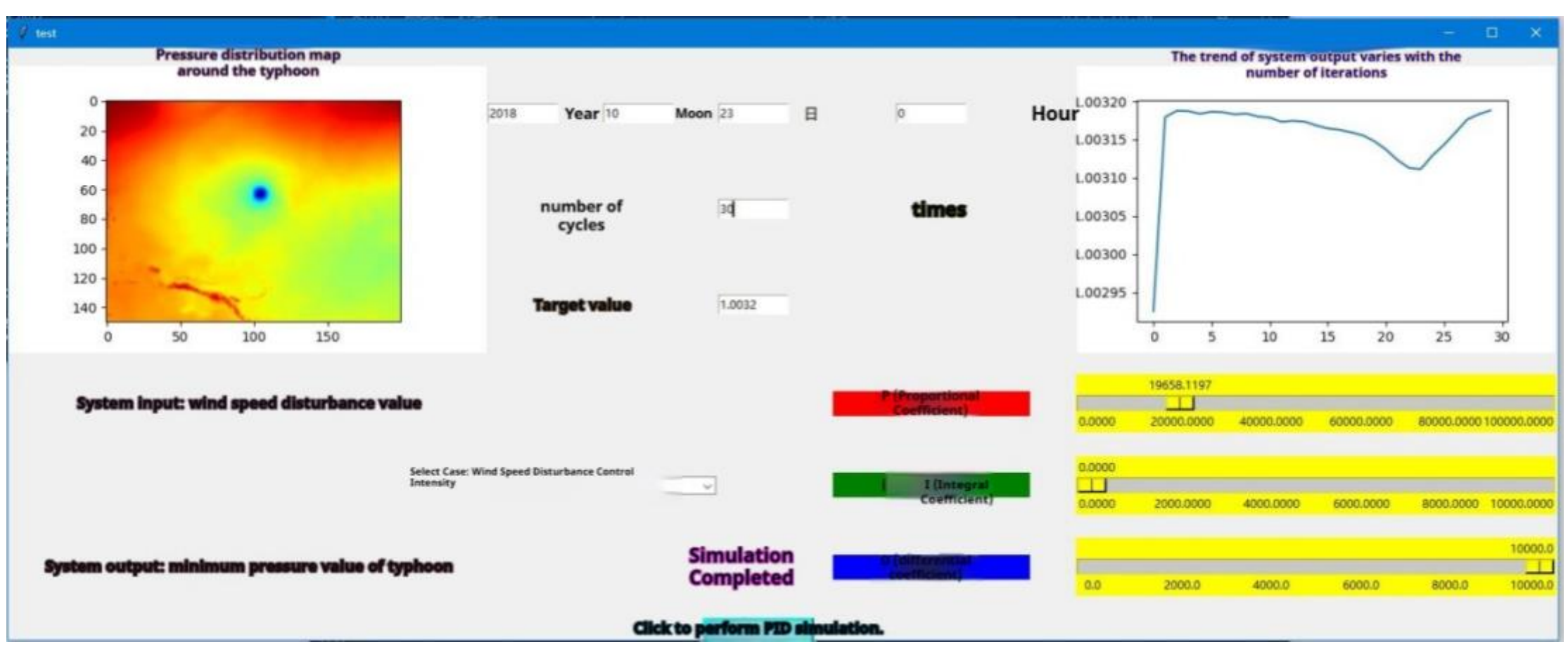

Figure 18 Example of artificial control typhoon simulation system with Pangu fusion PID

## 6 Conclusion and Discussion

This study successfully constructs a simulation methodology testbed for typhoon sensitivity analysis with the Pangu Weather Large Model as the core prediction module and PID closed-loop experimental techniques. A complete technical framework using MATLAB and ONNX cross-platform integration is established. The testbed features modular design for prediction, perturbation input, typhoon modeling, and PID-controlled response testing, providing a quantitative perturbation–response evaluation platform on operational AI weather models.

Key findings are as follows. First, precision testing of the Pangu model shows high accuracy for typhoon-related meteorological fields, with pressure field error below 0.3%, wind speed error about 0.3%, and ocean temperature error below 1%, and its 1.4-second inference speed provides necessary support for real-time iterative experimentation. Second, the artificial perturbation input interface effectively responds to external forcings, and the typhoon quantitative modeling module based on local minimum pressure and polar coordinates accurately converts abstract track and intensity into scalar quantities suitable for perturbation–response analysis. Third, the SISO test system demonstrates a clear feasible perturbation–response range, and PID parameter tuning follows traditional control theory rules, with distinct energy-scale response characteristics observed between velocity and thermal perturbation modes. Fourth, comparison with Japan's MS8 program shows that this testbed realizes perturbation–response analysis on a large-scale operational AI weather forecasting model with actual typhoon prediction capabilities, providing an experimental framework that goes beyond MS8's current verification exclusively on the idealized Lorenz 63 model.

Current limitations include: reliance on SISO qualitative testing with a limited set of perturbation inputs; use of traditional linear PID techniques, which face challenges in strongly nonlinear regimes; and test parameters derived from extreme super typhoon cases and specific perturbation configurations, limiting direct generalization to other scenarios.

Compared with previous studies using 4DVar in the WRF model, this work has its own advantages and disadvantages. The advantage of 4DVar is that it can consider nonlinearity, which is suitable for strong nonlinear high-dimensional chaotic systems such as typhoons. At the same time, 4DVar can be used in ensembles and belongs to optimal control methods and its mathematical expression is close to model predictive control, which is also the basis for EnMPC to integrate ensemble data assimilation and model predictive control. However, the disadvantage is that the perturbations obtained by back-stepping solution carry spatial structure, which is difficult to achieve in reality, and it is also difficult to test the sensitivity of individual variables, while the solution speed is slow. The advantage of this work on the one hand lies in the fast reasoning speed of AI models, which is suitable for conducting a large number of sensitivity experiments. On the other hand, PID control can test the system's response to individual variables, which is helpful for finding tools that might achieve perturbations. However, its disadvantage is also obvious. PID control is simple and mainly applicable to linear systems. For strong nonlinear systems such as typhoons, it cannot achieve optimal control and can only test qualitative responses or sensitivity analysis. It is necessary to shift to modern control methods such as MPC.

Future directions include: expanding to MIMO test systems with diverse perturbation forcings and multiple output indicators; integrating advanced nonlinear analysis methods and eventually EnMPC-based experimentation; designing perturbation configurations with lower energy scales for application-oriented studies; and adapting the testbed for sensitivity analysis of other extreme weather events such as hurricanes and rainstorms.

A direct comparison with Japan's Moonshot Goal 8 (MS8) program sharply illuminates the advancement offered by this testbed. MS8 has developed a sophisticated Ensemble Model Predictive Control (EnMPC) framework for weather intervention; however, its validation has been confined entirely to the idealized three-variable Lorenz-63 model, a system that, while chaotic, lacks the spatial dimensionality, physical realism, and operational forecasting capability needed to assess real-world perturbation responses for a mesoscale phenomenon such as a typhoon. In contrast, the testbed reported here accomplishes perturbation–response experiments on an operational-grade AI weather prediction model (Pangu) that has demonstrated verification skill competitive with ECMWF's IFS. Quantitatively, the testbed has delivered the first set of closed-loop output-tracking results for an actual typhoon case: it systematically mapped the track angle response to prescribed velocity perturbations converging to ~43 $m \cdot s^{-1}$, and characterized the central pressure response to target values of 1.0032 bar with converged disturbance magnitudes of ~16.5 $m \cdot s^{-1}$, among other SISO demonstrations. These numerical outcomes, obtained under realistic meteorological initial conditions, provide concrete data characterizing the perturbation–response relationships of an AI-represented typhoon and reveal the energy-scale characteristics of different forcing modes. By moving from an abstract, low-order dynamical system to a full-scale, high-resolution AI weather model, this testbed bridges a critical gap between idealized sensitivity theory and the practical assessment of typhoon response to environmental perturbations, laying a data-grounded foundation for subsequent expansion toward MIMO architectures and robust perturbation analysis strategies.

**Data availability**

Data will be made available on request.

# REFERENCES

Alam, S. R., Gila, M., Klein, M., Martinasso, M., and Schulthess, T. C., 2023: Versatile software-defined HPC and cloud clusters on ALPS supercomputer for diverse workflows. *The International Journal of High Performance Computing Applications*, **37**(3–4), 288–305, https://doi.org/10.1177/10943420231167811.

Bauer, P., Thorpe, A., and Brunet, G., 2015: The quiet revolution of numerical weather prediction. *Nature*, **525**(7567), 47–55, https://doi.org/10.1038/nature14956.

Bauer, P., Dueben, P. D., Hoefler, T., Quintino, T., Schulthess, T. C., and Wedi, N. P., 2021: The digital revolution of Earth-system science. *Nature Computational Science*, **1**(2), 104–113, https://doi.org/10.1038/s43588-021-00023-0.

Bauer, P., Dueben, P., Chantry, M., Doblas-Reyes, F., Hoefler, T., McGovern, A., and Stevens, B., 2023: Deep learning and a changing economy in weather and climate prediction. *Nature Reviews Earth & Environment*, **4**(8), 507–509, https://doi.org/10.1038/s43017-023-00468-z.

Ben Bouallègue, Z., Weyn, J. A., Clare, M. C. A., Dramsch, J., Dueben, P., and Chantry, M., 2024: Improving medium-range ensemble weather forecasts with hierarchical ensemble transformers. *Artificial Intelligence for the Earth Systems*, **3**(1), e230027, https://doi.org/10.1175/aies-d-23-0027.1.

Ben Bouallègue, Z., Clare, M. C. A., Magnusson, L., Gascón, E., Maier-Gerber, M., Janoušek, M., Rodwell, M., Pinault, F., Dramsch, J. S., Lang, S. T. K., Raoult, B., Rabier, F., Chevallier, M., Sandu, I., Dueben, P., Chantry, M., and Pappenberger, F., 2024: The rise of data-driven weather forecasting: A first statistical assessment of machine learning-based weather forecasts in an operational-like context. *Bulletin of the American Meteorological Society*, **105**, E864–E883, https://doi.org/10.1175/BAMS-D-23-0162.1.

Bi, K., Xie, L., Zhang, H., Chen, X., Gu, X., and Tian, Q., 2022: Pangu-weather: A 3D high-resolution model for fast and accurate global weather forecast. *arXiv preprint*

*arXiv:2211.02556*, https://doi.org/10.48550/arXiv.2211.02556.

Bi, K., Xie, L., Zhang, H., et al., 2023: Accurate medium-range global weather forecasting with 3D neural networks. *Nature*, **619**, 533–538, https://doi.org/10.1038/s41586-023-06185-3.

Bonavita, M., and Laloyaux, P., 2020: Machine learning for model error inference and correction. *Journal of Advances in Modeling Earth Systems*, **12**(12), e2020MS002232, https://doi.org/10.1029/2020MS002232.

Carrassi, A., Bocquet, M., Bertino, L., and Evensen, G., 2018: Data assimilation in the geosciences: An overview of methods, issues, and perspectives. *Wiley Interdisciplinary Reviews: Climate Change*, **9**(5), e535, https://doi.org/10.1002/wcc.535.

Cheon, M., and Mun, C., 2023: The climate of innovation: AI's growing influence in weather prediction patents and its future prospects. *Sustainability*, **15**(24), 16681, https://doi.org/10.3390/su152416681.

Fan, P. Y., Yang, J., Zhang, Z. P., Zang, N. H., Li, Y. F., and Feng, G. L., 2023: Summer precipitation prediction in eastern China based on machine learning. *Climate Dynamics*, **60**(9–10), 2645–2663, https://doi.org/10.1007/s00382-022-06464-1.

Ham, Y. G., Kim, J. H., and Luo, J. J., 2019: Deep learning for multi-year ENSO forecasts. *Nature*, **573**(7775), 568–572, https://doi.org/10.1038/s41586-019-1559-7.

Ham, Y. G., Kim, J. H., Kim, E. S., and On, K. W., 2021: Unified deep learning model for El Niño/Southern Oscillation forecasts by incorporating seasonality in climate data. *Science Bulletin*, **66**(13), 1358–1366, https://doi.org/10.1016/j.scib.2021.03.009.

Hess, P., Drüke, M., Petri, S., Strnad, F. M., and Boers, N., 2022: Physically constrained generative adversarial networks for improving precipitation fields from Earth system models. *Nature Machine Intelligence*, **4**(10), 828–839, https://doi.org/10.1038/s42256-022-00540-1.

Hipel, K. W., Jamshidi, M. M., Tien, J. M., and White III, C. C., 2007: The future of systems, man, and cybernetics: Application domains and research methods. *IEEE Transactions on Systems, Man, and Cybernetics*, **37**, 726–743, https://doi.org/10.1109/tsmcc.2007.900671.

Kim, H., Ham, Y. G., Joo, Y. S., and Son, S. W., 2021: Deep learning for bias correction of MJO prediction. *Nature Communications*, **12**, 3087, https://doi.org/10.1038/s41467-021-23406-3.

LeCun, Y., Bengio, Y., and Hinton, G., 2015: Deep learning. *Nature*, **521**, 436–444, https://doi.org/10.1038/nature14539.

Lei, H., Lei, W., and Zeng, Q., 2012: Cybernetics in the artificial weather modification. I: Direct and inverse or optimal controlling problem for precipitation enhancement operation. *Climatic and Environmental Research* (in Chinese), **17**(6), 968–978,

https://doi.org/10.3878/j.issn.1006-9585.2012.06.33.

Li, P. Y., Yu, Y., Huang, D. N., Wang, Z. H., and Sharma, A., 2023: Regional heatwave prediction using graph neural network and weather station data. *Geophysical Research Letters*, **50**(7), e2023GL103405, https://doi.org/10.1029/2023GL103405.

Ling, F. H., Luo, J. J., Li, Y., Tang, T., Bai, L., Ouyang, W. L., and Yamagata, T., 2022: Multi-task machine learning improves multi-seasonal prediction of the Indian Ocean Dipole. *Nature Communications*, **13**(1), 7681, https://doi.org/10.1038/s41467-022-35412-0.

Mu, B., Qin, B., and Yuan, S. J., 2021: ENSO-ASC 1.0.0: ENSO deep learning forecast model with a multivariate air-sea coupler. *Geoscientific Model Development*, **14**(11), 6977–6999, https://doi.org/10.5194/gmd-14-6977-2021.

Pan, B. X., Anderson, G. J., Goncalves, A., Lucas, D. D., Bonfils, C. J. W., and Lee, J., 2022: Improving seasonal forecast using probabilistic deep learning. *Journal of Advances in Modeling Earth Systems*, **14**, e2021MS002766, https://doi.org/10.1029/2021MS002766.

Ravuri, S., and Coauthors, 2021: Skilful precipitation nowcasting using deep generative models of radar. *Nature*, **597**(7878), 672–677, https://doi.org/10.1038/s41586-021-03854-z.

Shen, X., Wang, J., Li, Z., Chen, D., and Gong, J., 2020, Research and operational development of numerical weather prediction in China. *Journal of Meteorological Research*, **34**, 675–698, https://doi.org/10.1007/s13351-020-9847-6.

Song, J., Xue, H., Bao, X., Wu, D., Chai, F., Shi, L., Yao, Z., Wang, Y., Nan, F. and Wan, K., 2011, A spectral mixture model analysis of the Kuroshio variability and the water exchange between the Kuroshio and the East China Sea. *Journal of Oceanology and Limnology*, **29**, 446–459, https://doi.org/10.1007/s00343-011-0114-7.

Song, J., Guo, J., Li, J., Mu, L., Liu, Y., Wang, G., Li, Y. and Li, H., 2017: Definition of water exchange zone between the Bohai Sea and Yellow Sea and the effect of winter gale on it. *Acta Oceanologica Sinica*, **36**, 17–25, https://doi.org/10.1007/s13131-017-0989-z.

Tunstel, E., Cobo, M. J., Herrera-Viedma, E., Rudas, I. J., Filev, D., Trajkovic, L., Chen, C. L. P., Pedrycz, W., Smith, M. H., and Kozma, R., 2021**:** Systems science and engineering research in the context of systems, man, and cybernetics: Recollection, trends, and future directions. *IEEE Transactions on Systems, Man, and Cybernetics: Systems*, **51**, 5–21, https://doi.org/10.1109/tsmc.2020.3043192.